\documentclass[preprint,showpacs,tightenlines,preprintnumbers,amsmath,amssymb]{revtex4}

\usepackage[dvips]{graphicx}

\newcommand{\br}{\vec{r}}
\def\be{\begin{equation}}
\def\ee{\end{equation}}
\def\bea{\begin{eqnarray}}
\def\eea{\end{eqnarray}}

\begin{document}

\title{Isospin asymmetry in the continuum of the A=14 mirror nuclei}

\author{H.  Sagawa}

\affiliation{Center for Mathematical Sciences,  University of Aizu\\
Aizu-Wakamatsu, Fukushima 965-8560,  Japan\\
E-mail: sagawa@u-aizu.ac.jp}


\begin{abstract}
We study the isospin asymmetry in the isoscalar (IS) excitations
 in the mirror nuclei $^{14}$O and
$^{14}$C by using the Hartree-Fock(HF)+random phase approximation (RPA)
 linear response function theory with a Skyrme interaction to take
 into account the continuum effect properly.
The asymmetry in the IS  monopole, dipole  responses is  
pointed out in the continuum near the particle threshold with respect to 
 the excitation energy and the sum rule strength. On the other hand,  
no clear sign of the asymmetry is found in 
 the giant resonance (GR) region.  In the quadrupole case, 
  the calculated strengths of the mirror nuclei show  almost the
same energy dependence from the threshold to the GR region.
It is found  that the transition densities of the monopole response 
show an extended halo structure near the threshold, while those of GR region 
show a typical  radial dependence of the compressional
 collective mode without any halo effect.
Contrary to the transition densities of  the monopole response,  
 those of quadrupole response do not 
   show any sign of the extended feature of wave functions  neither near the 
threshold nor the GR energy region.
Calculated strength distributions  of the IS multipole states   
are compared with  recent experimental
data obtained by the multipole decomposition analysis of 
$\alpha $ inelastic scattering on  $^{14}$O. 

\end{abstract}

\maketitle

\section{Introduction}
A fundamental feature of nuclear structure is associated with 
 a basic symmetry between neutrons and protons in the nuclear 
interactions.  The existence of a general symmetry between $np,nn$ and 
$pp$ interactions was first observed in the low$-$energy nucleon$-$nucleon
scattering data.  The charge independence of the interaction  implies the 
conservation of the isospin symmetry in nuclei \cite{Heisenberg32,BM69}.  
The charge symmetry 
 assumption, based on the equality of the $nn$ and  
$pp$ interactions , might  give almost the same excitation spectra  in 
nuclei of the isospin partners.
  The isospin symmetry, however, is violated by the electromagnetic 
 interaction. Additional symmetry$-$breaking effects
  arise from the neutron$-$proton mass 
 difference, and small charge symmetry and independence breaking forces 
 in the strong interactions.  

 The isospin impurity was studied by
using microscopic models in refs.  \cite{HS93,HGS95} in relation with 
Gamow$-$Teller and Fermi $\beta$ decays. It was pointed out that the isospin 
mixing increases rapidly as a function of proton number and becomes few \% 
in the ground state of $^{100}$Sn.  
However it will be less than 0.1\% in light nuclei with the mass less than 
A=16  and thus the isospin can be considered  to be  a good and useful 
quantum number in  the light nuclei. 
  For example, in the A=13 mirror nuclei $^{13}$N and 
 $^{13}$C with the isospin T=1/2,  all states in the  spectra show 
 one to
 one correspondence up to the excitation energy  Ex=10MeV within  
  few hundreds keV accuracy. 

 In heavier nuclei, it was expected that the isospin symmetry 
might be of little significance because of very strong Coulomb field.
  However, the experimental observation of the isobaric analog states (IAS)
 in medium heavy nuclei shows that the isospin is well preserved even 
 under the influence of the strong Coulomb field \cite{Fox64}.  
 In light nuclei, 
 the examples of the large isospin asymmetry were found in the first excited 
J$^{\pi}$=1/2$^+$ 
 states  of A=13 nuclei, 
$^{13}$C and $^{13}$N,  and also in the J$^{\pi}$=1/2$^+$ 
 states  of A=17 nuclei $^{17}$O and $^{17}$F.
  This is called Thomas$-$Ehrman effect \cite{ET51}.
  This large asymmetry is considered mainly due to the reduction of the 
Coulomb energy of 
loosely$-$bound 
 low$-l$ state with small centrifugal barrier. 
 It should be noticed that
 a large anomaly was also found in the isoscalar (IS)
  magnetic moments in A=9 mirror 
nuclei experimentally \cite{Matsuta95} and discussed in relation with the Thomas$-$Ehrman effect \cite{Utsuno04}. 

 The Coulomb energy might be reduced 
 not only for the loosely bound states but also for the unbound resonance 
and continuum 
 states.  In this paper, we will study the continuum states
  in the T=1 mirror nuclei
   $^{14}$O and  $^{14}$C in order to extract the isospin asymmetry 
in the spectra.  
The A=14 isospin partners are particularly interesting 
 because of a large difference in the threshold energies of
 the mirror nuclei.
Namely, 
the neutron threshold energy of $^{14}$C is  S$_n$=8.18MeV,  while the 
proton threshold energy in  $^{14}$O
 is S$_p$=4.63MeV.   
This large energy difference between the 
two nuclei  may give rise to a substantial  effect on the 
 low multipole strength distributions with 
J$^{\pi}$=0$^{+}$ and 1$^{-}$, especially near the threshold. 
The lowest 1$^-$ state in   $^{14}$O  has been obtained  much  attention
due to its importance on the astrophysical CNO cycle in the stars 
\cite{Moto91}.
The main aim of this paper is to study the isospin asymmetry 
of the continuum response functions in the A=14 mirror nuclei both near the 
threshold and giant resonance region.   To this end,  
 we use the  
  Hartree$-$Fock (HF) $+$ random phase approximation (RPA)
 continuum response function with a Skyrme interaction 
to calculate the IS  strength distributions 
 of low multipole states in   $^{14}$O and  $^ {14}$C. 
 The two-body spin-orbit and Coulomb residual interactions are not 
 included in the RPA response. 
   This paper is organized as follows.
A brief  review  of the 
RPA response function is given in  Section II.  Calculated results are 
shown in Section III and compared with available experimental data.
 A summary is given in Section IV.

\section{Continuum strength in $^{14}$O and  $^{14}$C}
The RPA linear response theory is based on 
the time$-$dependent HF (TDHF) theory \cite{BT75}.  
The RPA Green's function  $G^{{\rm RPA}}$ is obtained as a small 
amplitude approximation of TDHF and expressed as
\begin{eqnarray}
G^{{\rm RPA}} & = & G^{(0)} + G^{(0)}\frac{\delta v}{\delta \rho }
   G^{{\rm RPA}}\nonumber\\
              & = & (1- G^{(0)}\frac{\delta v}{\delta \rho }
          )^{-1}G^{(0)}.
\label{eq:b5}
\end{eqnarray}
where $\frac{\delta v}{\delta \rho }$ is the residual particle$-$hole 
interaction.  The unperturbed Green's function  $G^{(0)}$ is defined as 
\begin{eqnarray}
G^{(0)}(\br, {\br}':\omega) = \sum _{i \in occupied}{{\varphi}_i}^{*}
                          (\br)\left<\br \left| \frac{1}{ \omega +i\eta
            -h_0 + {\epsilon}_i} -   \frac{1}{\omega
                            - i\eta + h_0 - {\epsilon}_i  } \right| {\br}'
       \right> \varphi _i ({\br}') .   
\label{eq:b7}
\end{eqnarray}
where $\varphi_i$ and ${\epsilon}_i$ are the eigenfunction and the
eigenenergy of the HF hamiltonian $h_0$.  
The operator equation in the r.h.s. of (\ref{eq:b7}) is nothing 
but the one-body 
Green's function in the coordinate space representation. We can use the 
standard 
technique to solve the Green's function taking into account the coupling to
the continuum ~\cite{LG76}.  

The transition strength 
$S(\omega)$ for the states above the threshold
 can be obtained from the imaginary part of $G^{{\rm RPA}}$ as
\begin{eqnarray}
S(\omega) &\equiv& \: \sum_{n} \mid <n\mid f^{\lambda , \tau}(\br )
 \mid 0> \mid ^{2} \delta (\omega -E_{n})  \nonumber \\
& =& \frac 1 \pi \int \int Im
   \left\{ f^{\lambda, \tau \dagger} ({\br }') G^{{\rm
     RPA}}(\br ,{\br }';\omega)f^{\lambda, \tau}(\br )\right\}d\vec{r}d\vec{r'}
\nonumber \\
\label{eq:b12}
\end{eqnarray}
where $f^{\lambda , \tau}(\br)$ is the transition operator with
multipolarity $\lambda $ and isospin $\tau $.    Below the  threshold, 
the strength is calculated 
from the residue of the poles in the real part of the response
function.
The transition
operators are given by
\begin{eqnarray}
\label{eq:opemnis}
f^{\lambda=0, \tau=0} \: &=& \: \sum_{i=1}^{A} r^{2}_{i}
\frac{1}{\sqrt{4\pi }}
\qquad \mbox{for isoscalar monopole strength},  
\end{eqnarray}
\begin{eqnarray}
\label{eq:opeqpis}
f_{\mu}^{\lambda=2, \tau=0} \: &=&  \: \sum_{i=1}^{A} r^{2}_{i} Y_{2\mu}(\hat{r}_{i})
\qquad \mbox{for isoscalar quadrupole strength}.
\end{eqnarray}
The IS dipole operator
$\sum_{i} r_{i} Y_{1\mu}(\hat{r}_{i})$ excites  the 
spurious state corresponding to the center-of-mass motion. 
Thus, we consider the next order term in the 
expansion of the spherical Bessel function 
$j_{\lambda=1}(qr)$ for $qr \! \ll \! 1$,
\begin{equation}
\label{eq:opedpis}
f_{\mu}^{\lambda=1, \tau=0} \: =  \: \sum_{i=1}^{A} r^{3}_{i} Y_{1\mu}(\hat{r}_{i})
\qquad \mbox{for isoscalar compression  dipole  strength}.\\
\end{equation}
The modified IS dipole operator is also defined as 
\begin{equation}
\label{eq:opedpism}
f_{\mu}^{\lambda=1, \tau=0} \: =  \: \sum_{i=1}^{A} (r^{3}_{i}-\eta r_i) Y_{1\mu}(\hat{r}_{i})
\end{equation}
where $\eta =3<r^2>/5$ subtracts the spurious 
component from the operator (\ref{eq:opedpis}) \cite{Harakh81,HSZ98}.
 
The energy weighted sum rule (EWSR) is defined to be
\begin{equation}
\label{eq:ewsr}
S_{\lambda}=\sum_{n,\mu}|<n|f^{\lambda}_{\mu}|0>|^2
        =\frac{1}{2}\sum_{\mu}<0|[f^{\lambda *}_{\mu},[H,f^{\lambda}_{\mu}]]|0>
\end{equation}
where $H$ is the hamiltonian for HF and RPA calculations.
For the IS excitations, the Skyrme two-body interactions do not give any 
contributions for the EWSR \cite{BT75}.  Thus the EWSR can be expressed 
 for the operators (4), (5) and (7) to be
\begin{eqnarray}
S_{\lambda=0}&=&\frac{\hbar^2}{2m}\frac{1}{\pi}A<r^2> \\
S_{\lambda=2}&=&\frac{\hbar^2}{2m}\frac{50}{4\pi}A<r^2> \\
S_{\lambda=1}&=&\frac{\hbar^2}{2m}\frac{3}{4\pi}A(11<r^4>-\frac{25}{3}<r^2>^2).
\end{eqnarray}

The transition density for an excited state, ~$\mid n\!>$~, 
\begin{equation}
\label{eq:td}
\delta \rho (\vec{r}) \: \equiv \: <n\mid \sum_{i} 
\delta(\vec{r}-\vec{r}_i) \mid 0>
\end{equation}
can be obtained from the RPA response, since the imaginary part of 
~$G^{RPA}(\vec{r} \: , \: \vec{r'} \: ; \, E_{n})$~ 
near the resonance  is
proportional to 
~$\delta \rho(\vec{r}) \, \delta \rho(\vec{r'})^{*}$~.
The reduced transition probability is calculated using 
the transition density (\ref{eq:td})   as
\begin{equation}
\label{eq:BEL}
B(\lambda, \tau : 0 \rightarrow n) \: = \: \sum_{\mu} \mid <n\mid
f^{\lambda ,  \tau}_{\mu} \mid 0> \mid ^2
=\: \sum_{\mu} \mid \int 
\delta \rho(\vec{r})\, f^{\lambda ,
  \tau}_{\mu}(\vec{r})\,
 d\vec{r} \mid ^2
\end{equation}
where we use an identity of the one-body transition operator
\begin{eqnarray}
f^{\lambda ,  \tau}_{\mu}= \int \sum_{i}  
\delta(\vec{r}-\vec{r}_i) f^{\lambda ,  \tau}_{\mu}(\vec{r})
d\vec{r} .                              \nonumber 
\end{eqnarray}
The radial transition density,~$\delta \rho_{\lambda}(r)$~, is defined by 
\begin{equation}
\delta \rho(\vec{r}) \: \equiv \:\delta  \rho_{\lambda}(r) \: 
Y^{*}_{\lambda \: \mu}(\hat{r}) \qquad .
\label{eq:td2}
\end{equation}

\section{Discussions}
The Skyrme interaction SkM$^*$ is used to calculate
   HF potentials and 
RPA response functions.   The HF calculations  are performed with 
the filling approximation putting all nucleons in the HF orbits 
from  the bottom of the mean field potentials. In $^{14}$O, the 
proton number Z=8 corresponds to the closed shell while 
the neutron number N=8 in  $^{14}$C is the closed shell.
Moreover, the energy difference between the $1p_{3/2}$ and $1p_{1/2}$ 
states for neutrons in $^{14}$O and protons  in  $^{14}$C are
 more than 6MeV in the HF calculations.  Thus, the pairing correlations
might not play any important role in the RPA response in these nuclei 
\cite{Khan}.  

The Skyrme 
interaction SkM$^*$ gives  the neutron separation energy 
S$_n$=8.95MeV in  
 $^{14}$C and the proton separation energy S$_p$=5.70MeV in  $^{14}$O.  
These values are somewhat larger than the empirical ones 
S$_n$(exp)=8.18MeV in  $^{14}$C and  S$_p$(exp)=4.63MeV in $^{14}$O.
In order to perform  quantitative study of the threshold effect 
in the response, we enlarge the spin-orbit strength of the  SkM$^*$ 
interaction to be $W_o$=150MeV$\cdot$fm$^{5}$  in  $^{14}$C and 
$W_o$=155MeV$\cdot$fm$^{5}$  in  $^{14}$O from the
 original value $W_o$=130MeV$\cdot$fm$^{5}$.  These  modifications give 
 S$_n$=8.11MeV in $^{14}$C and  S$_p$=4.66MeV in $^{14}$O which are 
very close to the experimental ones.

The existing Skyrme parameters do not reproduce the empirical separation 
energies of the two nuclei  $^{14}$C and $^{14}$O with the  accuracy
which is good enough to study the threshold behavior of excitations
in loosely-bound nuclei.   It has been known that the particle-vibration 
coupling has a substantial effect on the single-particle energy 
, especially around the Fermi surface \cite{Colo}. For more sophisticated 
calculations, one should make an effort to readjust the Skyrme 
parameters including the particle-vibration coupling. 
Since it is not the aim of the present study to readjust all the 
Skyrme parameters, we take a conventional approach to obtain
 the empirical separation energies of A=14 nuclei.

\begin{figure}[h]
\includegraphics[width=5in,clip]{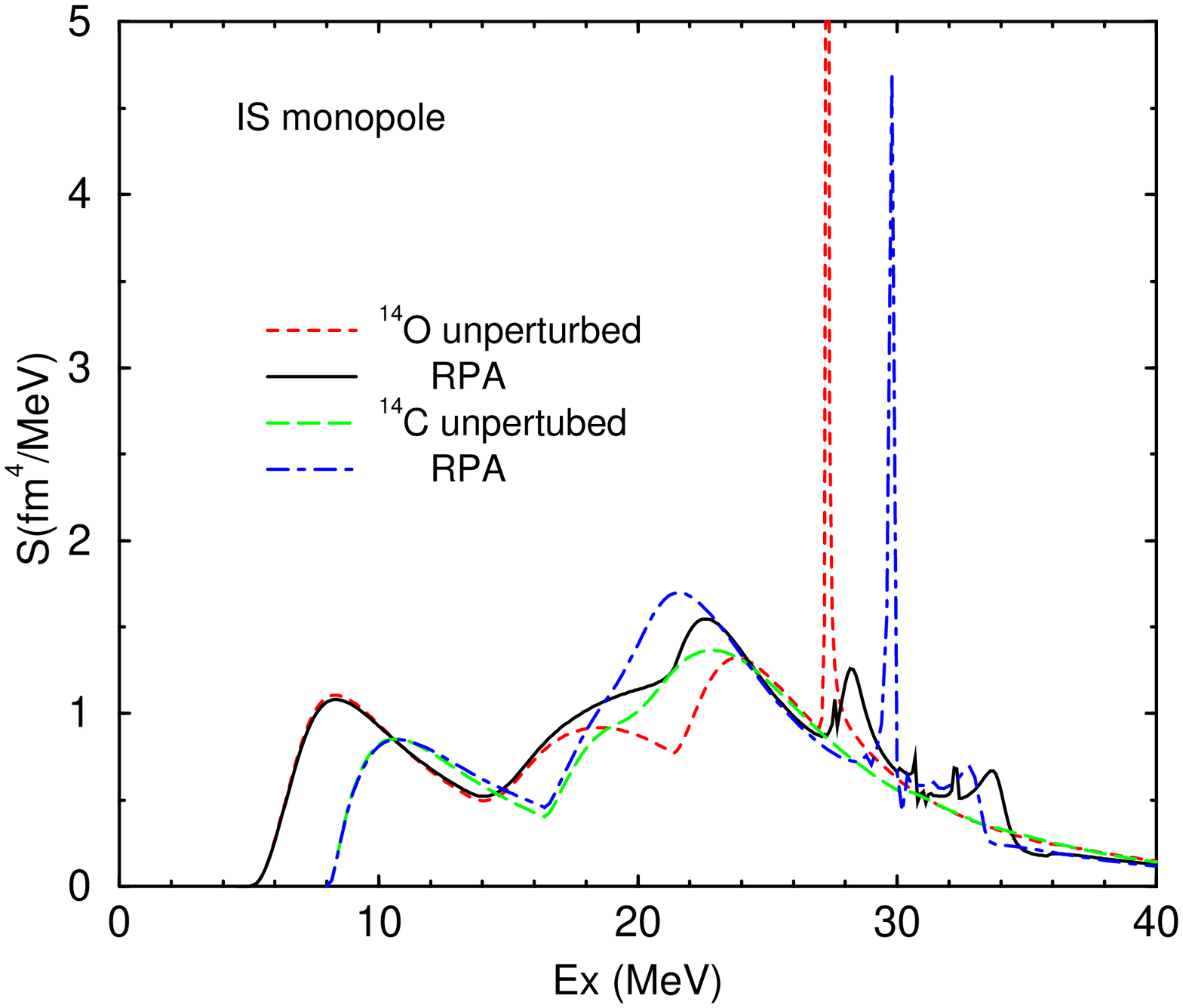}
\caption{\label{fig1} (Color online) 
Unperturbed and RPA IS monopole response $S$ in  
 $^{14}$C and $^{14}$O calculated by
the self-consistent  response function theory with the 
Skyrme interaction SkM$^*$. Unperturbed and RPA responses (\ref{eq:b12})
 are 
calculated by using the Green's functions (\ref{eq:b7}) and (\ref{eq:b5}), 
 respectively, with the operator (\ref{eq:opemnis}).
 See the text for details.}
 \end{figure}

The unperturbed 
and RPA 
isoscalar monopole responses of $^{14}$C and  $^{14}$O are given 
in Fig. \ref{fig1}.  There are two bound particle$-$hole 
excitations $\nu$, $\pi$(1s$_{1/2}\rightarrow$2s$_{1/2}$) ($\nu$ and $\pi$
 stand for neutron and proton, respectively, hereafter) 
at Ex$\sim$31MeV
in $^{14}$C, 
while  the proton particle state of the
 $\pi$(1s$_{1/2}\rightarrow$s$_{1/2}$) configuration 
is in the continuum in $^{14}$O and
 gives a sharp peak at  Ex$\sim$27.3MeV 
in the unperturbed response.  The discrete states of the unperturbed response 
, which are not shown in Fig. \ref{fig1},  couple with the 
continuum in the RPA response and appears as a sharp peak at Ex$\sim$30MeV
in $^{14}$C and also a peak at Ex$\sim$28.5MeV in $^{14}$O.  
The unperturbed peak   at  Ex$\sim$27.3MeV in $^{14}$O is integrated in the
giant resonance peak in the RPA response.

\begin{figure}[p]
\includegraphics[width=8cm,clip]{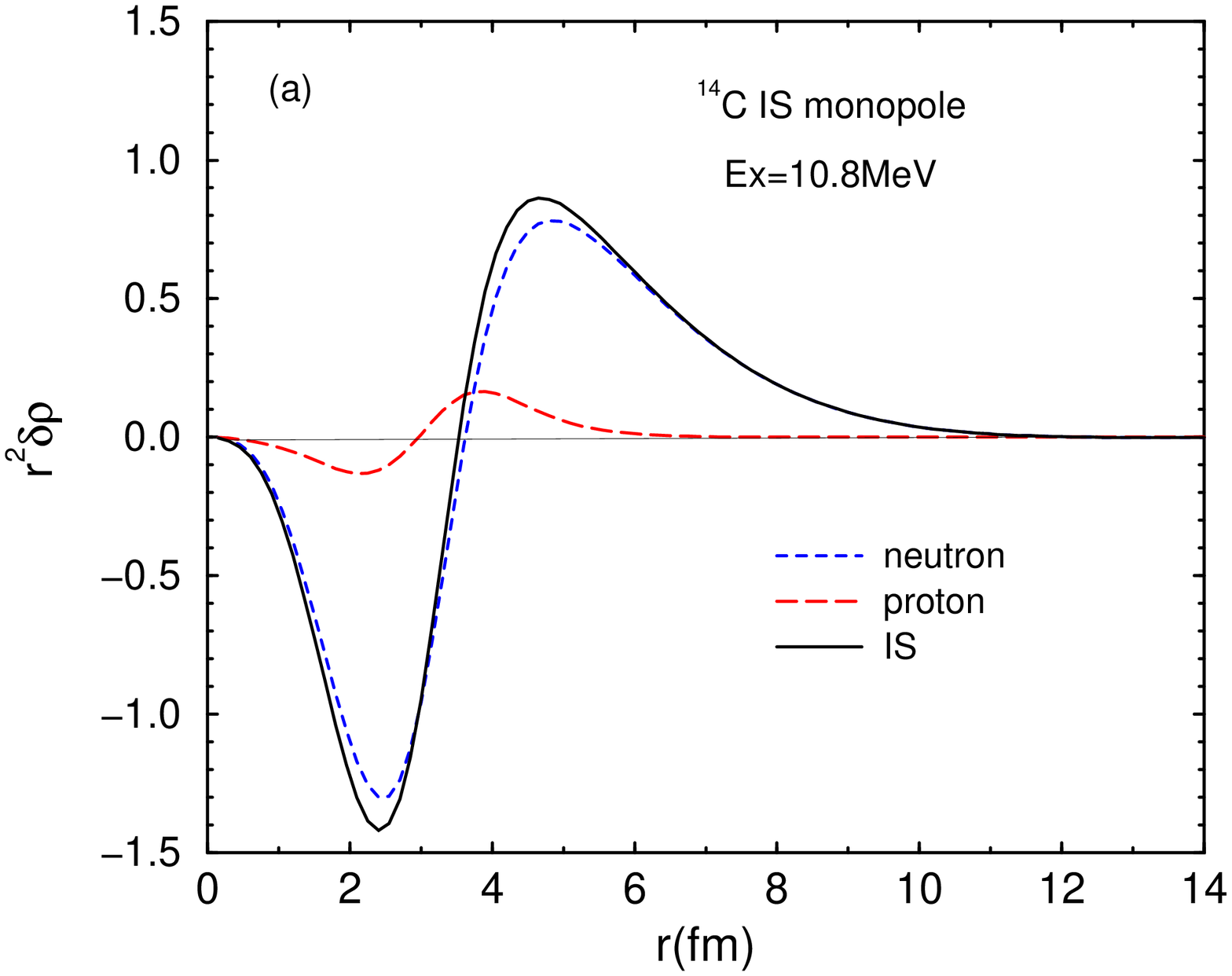}
\includegraphics[width=8cm,clip]{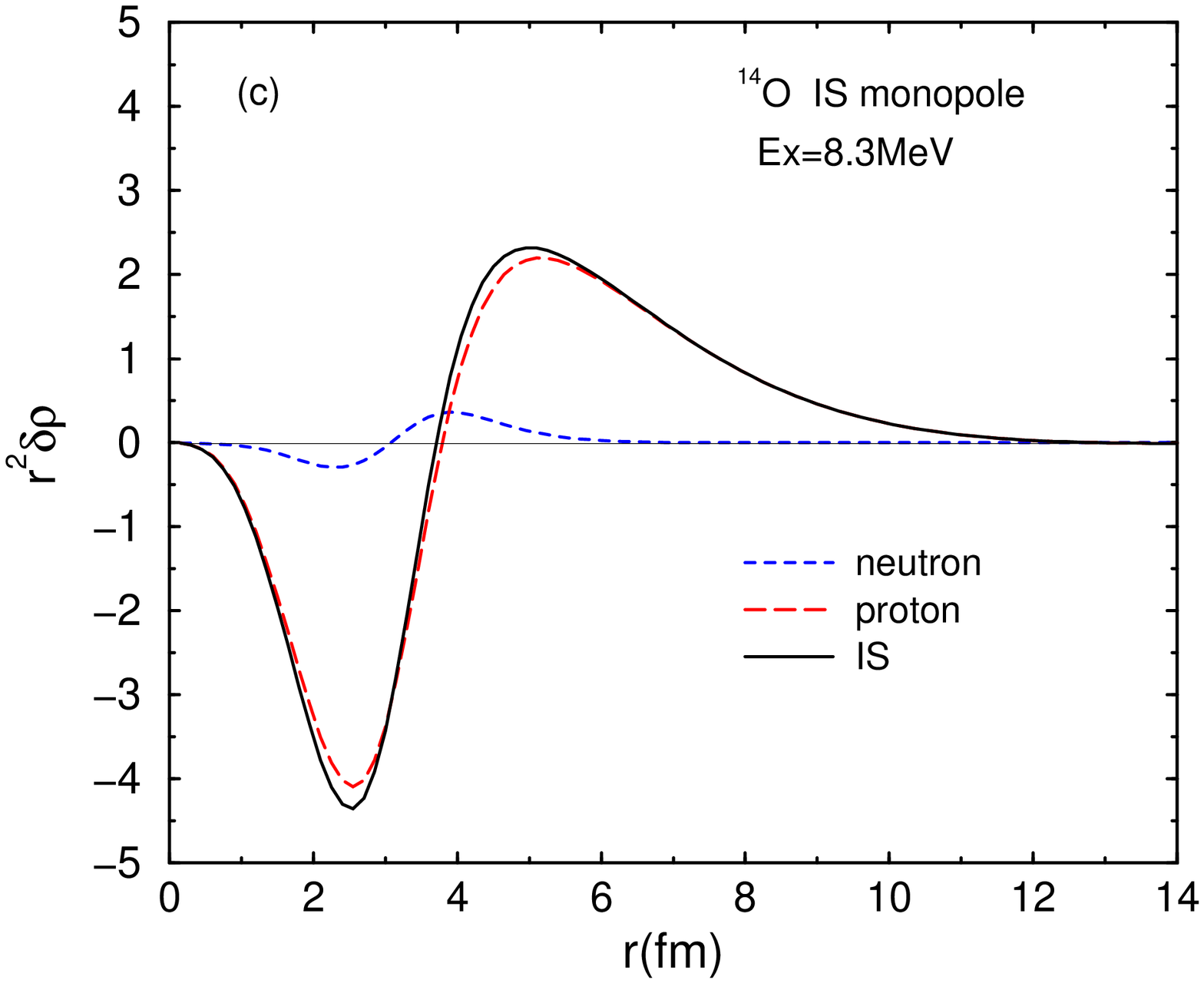}
\includegraphics[width=8cm,clip]{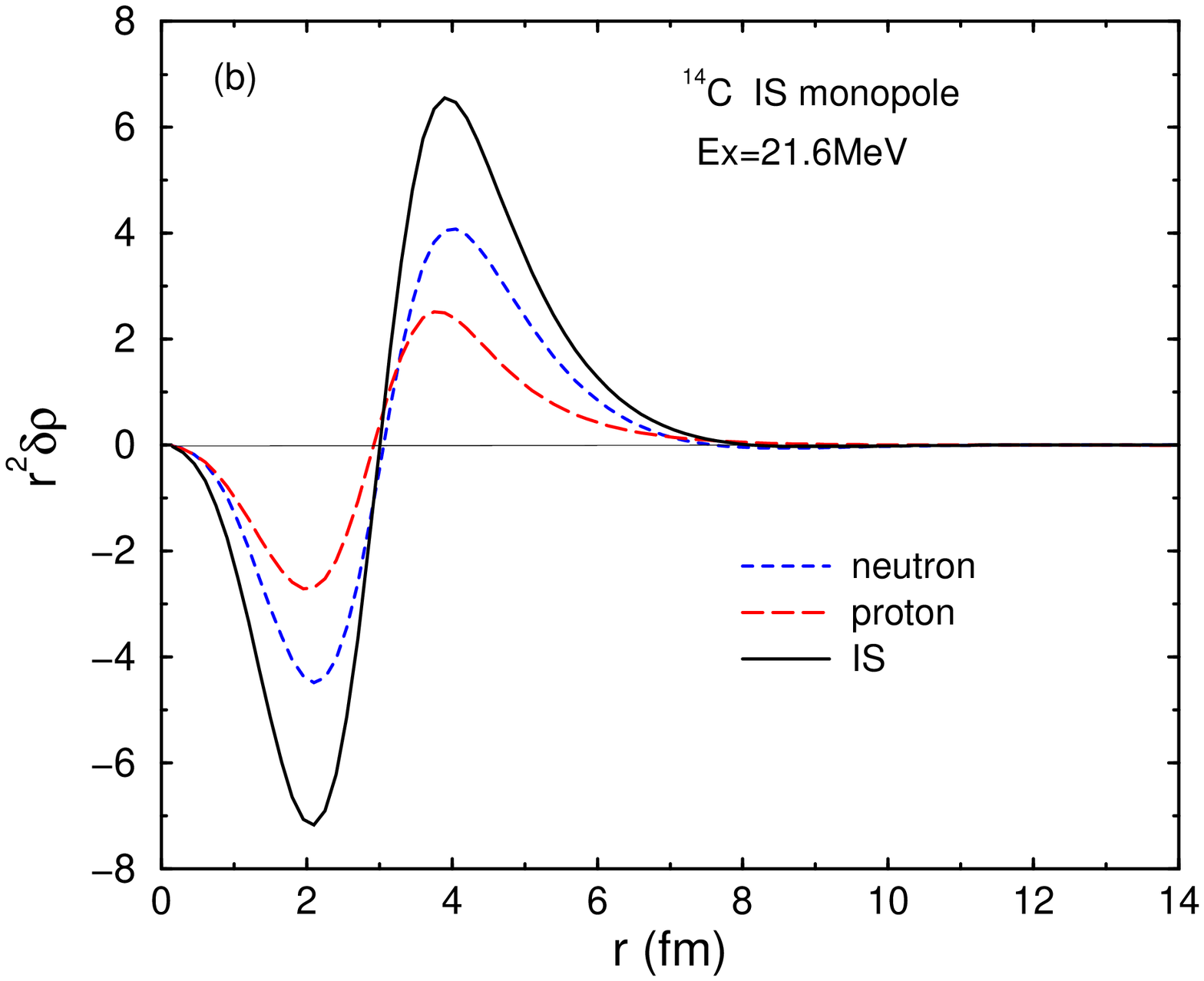}
\includegraphics[width=8cm,clip]{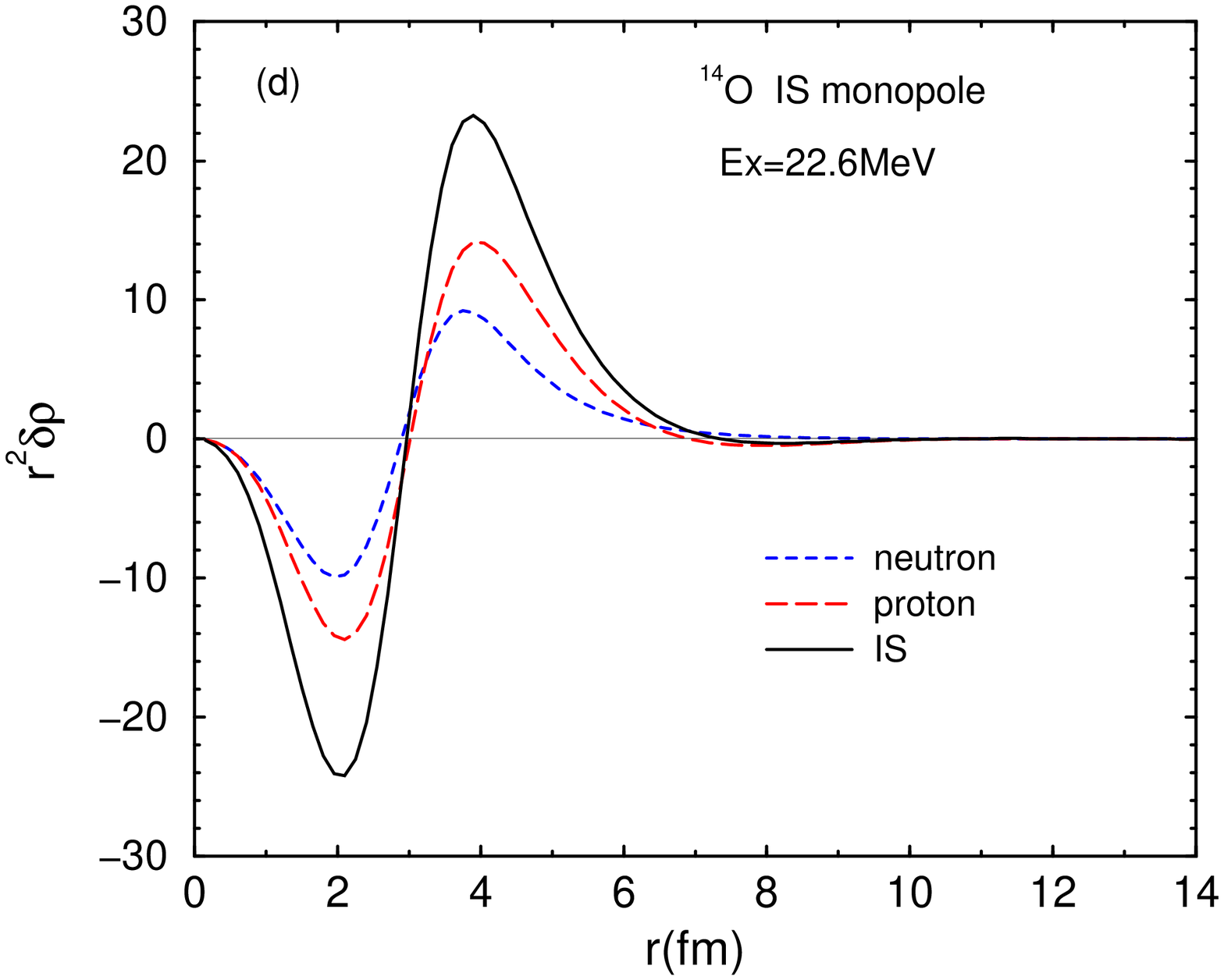}
\caption{\label{fig2} (Color online) 
Transition densities (\ref{eq:td}) of the IS monopole states in 
$^{14}$C and $^{14}$O; (a) and (c) at the threshold peaks,  and (b) and 
(d) at the GR peaks. The transition density is shown in an arbitrary unit.}
\end{figure}

The large threshold strength can be seen just above the proton threshold
in  $^{14}$O and also above the neutron threshold in $^{14}$C.
The energy of the threshold peak in $^{14}$O 
is more than one MeV lower that that in $^{14}$C due to the  lower proton 
separation energy S$_p$.
 One can see 
that the RPA correlations do  not make any appreciable changes on the 
unperturbed response around the threshold peaks as was 
pointed out in ref. \cite{HSZ97}.  
The monopole strength 
below 14MeV in  $^{14}$O exhausts 10.1\% of the energy weighted sum rule value
(EWSR), while the monopole strength below 14MeV in  $^{14}$C has 
only 8.1\% of the EWSR value. 
 The giant resonance peaks appear at Ex$\sim $22MeV
both in  $^{14}$O and $^{14}$C where RPA correlations enhance the monopole 
strength.  The integrated monopole strength up to GR region (Ex$\le$25MeV) is
similar in the two nuclei  49.6\% in   $^{14}$O and 48.5\% in $^{14}$C, 
respectively.  
The transition densities of the threshold peaks are shown in Figs. 
\ref{fig2} (a) for  $^{14}$C and (c) for  $^{14}$O. The dominant 
 contribution comes from the neutron excitation to the continuum in 
 $^{14}$C, while the proton continuum excitation has the major contribution 
in $^{14}$O.  The neutron density shows a halo-type long  tail 
 extended more than 10fm in the continuum excitation of $^{14}$C, 
    while the proton 
   density is rather compact and gives  a minor  contribution.
 The same halo effect appears in the proton transition density in $^{14}$O 
 near the threshold, while the neutron one is very small  and compact.
  Contrarily, the transition densities of GR region of $^{14}$C
 and  $^{14}$O
in Figs. 
\ref{fig2} (b) and (d), respectively,  
 have substantial contributions from 
both protons and neutrons.  The radial dependence
of the two contributions is very similar showing the peak and the node 
at almost the same radii.  The ratio between 
the proton and the neutron contributions around the peak at r$\sim$4fm is 
almost the same as that of the proton and the neutron numbers of
 each nucleus.  This is one of the typical feature of the collective IS 
 GR \cite{Sagawa02}.  
 In general, one can see in Fig. \ref{fig2} that the proton contribution 
to the transition densities in  $^{14}$O is almost identical to the neutron 
one in  $^{14}$C both in the radial dependence and  in the ratio to 
 another contribution; the ratio of proton to neutron contributions
 in  $^{14}$O is very close to that of neutron to proton ones in  $^{14}$C.  
The IS monopole strength in   $^{14}$O 
was observed recently by the multipole decomposition analysis 
of $\alpha $ inelastic scattering experiments \cite{Baba}. 
The IS peak is observed at Ex$\sim$7MeV and the integrated strength 
up to 13.8MeV exhausts about 8\% of the EWSR.  
These experimental data can be compared with 
 the calculated  peak  at 
 Ex$\sim$8MeV which exhausts 10\% of  the EWSR value.
The observed strength in the GR region is about 38\% of 
the EWSR up to Ex=25MeV in the  $\alpha$ inelastic scattering. 
This value is close to the integrated strength of RPA response
 up to Ex=25MeV, i.e.,  49.6\% 
of the EWSR.

\begin{figure}[p]
\includegraphics[width=5in,clip]{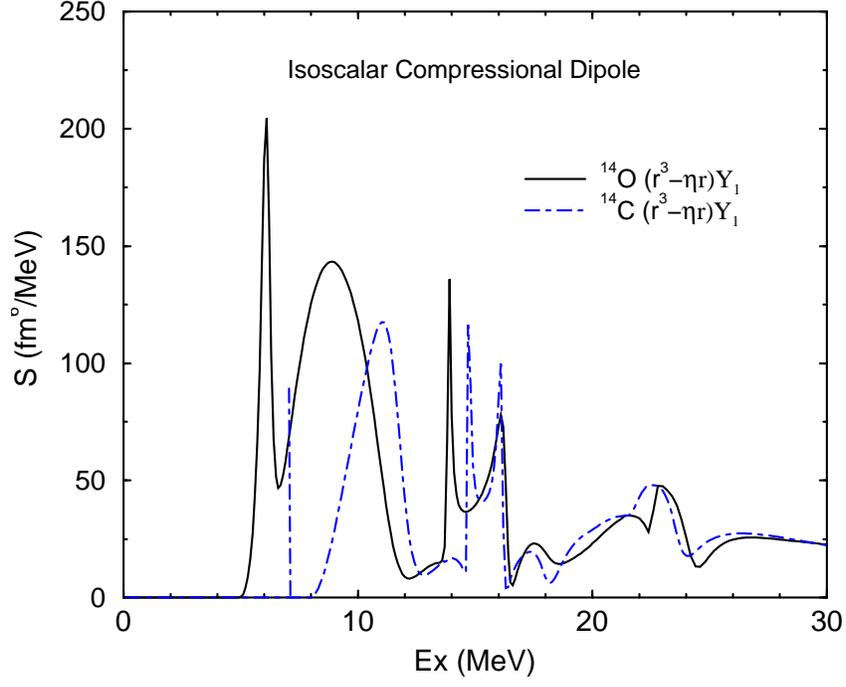}
\hspace{-2cm}
\caption{\label{fig3} (Color online) 
 IS compressional 
dipole response $S$ in  $^{14}$C and $^{14}$O calculated by
the self-consistent RPA response function theory with the 
Skyrme interaction SkM$^*$. The dipole operator (\ref{eq:opedpism}) 
is adopted in the calculations.  The dashed$-$dotted  line at Ex
=7.75MeV shows a discrete 1$^-$ state in $^{14}$C. See the text for details.}
\end{figure}

The isoscalar compressional 
 dipole  responses of $^{14}$C and  $^{14}$O are given 
in Fig. \ref{fig3}. In order to subtract accurately the spurious component, I 
took a small mesh size 0.075fm and the reference radius 15fm.
Then more than 95\% of the spurious center of mass motion was eliminated 
from the physical states \cite{HS02}. In  Fig. \ref{fig3}, 
we adopted the dipole  operator
 (\ref{eq:opedpism})  where the spurious component is further 
subtracted from the operator.   
The dipole response of $^{14}$O shows a sharp peak just above the 
threshold at Ex$\sim$6.1MeV 
   and  a broad bump peaked at Ex$\sim$9MeV. 
In $^{14}$C, a discrete 1$^-$ state is found at  Ex=7.1MeV
together with  a broad peak at  Ex$\sim$11MeV. 
 In the two nuclei, 
substantial strengths are found also above  Ex=15MeV.

 The observed
  lowest 
1$^-$ state is at Ex=5.17MeV in $^{14}$O and at 
  Ex=6.09MeV in $^{14}$C, respectively \cite{Firestone}. 
  It is remarkable  that the calculated energy difference  $\Delta $E=1MeV 
between the lowest 1$^-$ RPA states in the two nuclei  
 is  as large as 
the experimental one  $\Delta $E(exp)=0.92MeV,  although the observed 
  1$^-$ states    are found  at 
  somewhat lower excitation energies than the calculated 
ones.  The strength of the threshold 1$^-$ peak in  $^{14}$O has 2.5\% of 
the EWSR value, while the discrete lowest 1$^-$ state in $^{14}$C has only 
1.1\% of the EWSR strength.  The low energy peaks above the first excited 
states show also   
 large differences  both in  energy and in  transition strength. Namely, 
 the peak energy in  $^{14}$O is 8.9MeV, while that is shifted to about 2MeV 
higher in   $^{14}$C with  Ex(peak)$\sim$11MeV. 
 The integrated strength up to Ex=12MeV is 17.3\% of the EWSR in   $^{14}$O  which is much larger than that of the value in  
 $^{14}$C with  10.7\% of the EWSR value. On the other hand, 
 in the GR region, the two nuclei have almost 
the same transition strength, i.e.,  
 23\% of the EWSR between 12MeV $<$ Ex$<$25MeV both 
in $^{14}$O and  $^{14}$C.  

\begin{figure}[p]
\includegraphics[width=7cm,clip]{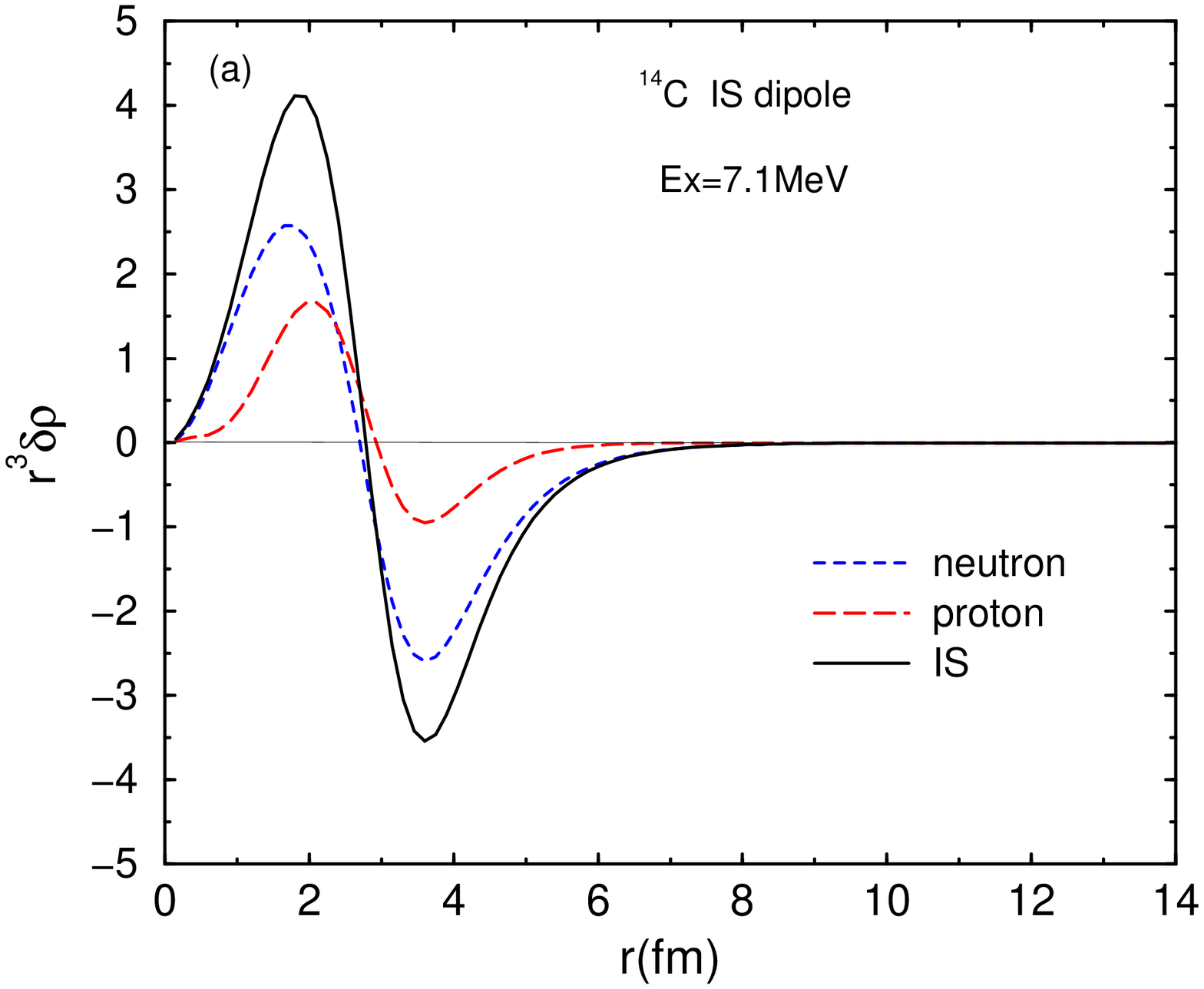}
\vspace{-2cm}
\includegraphics[width=7cm,clip]{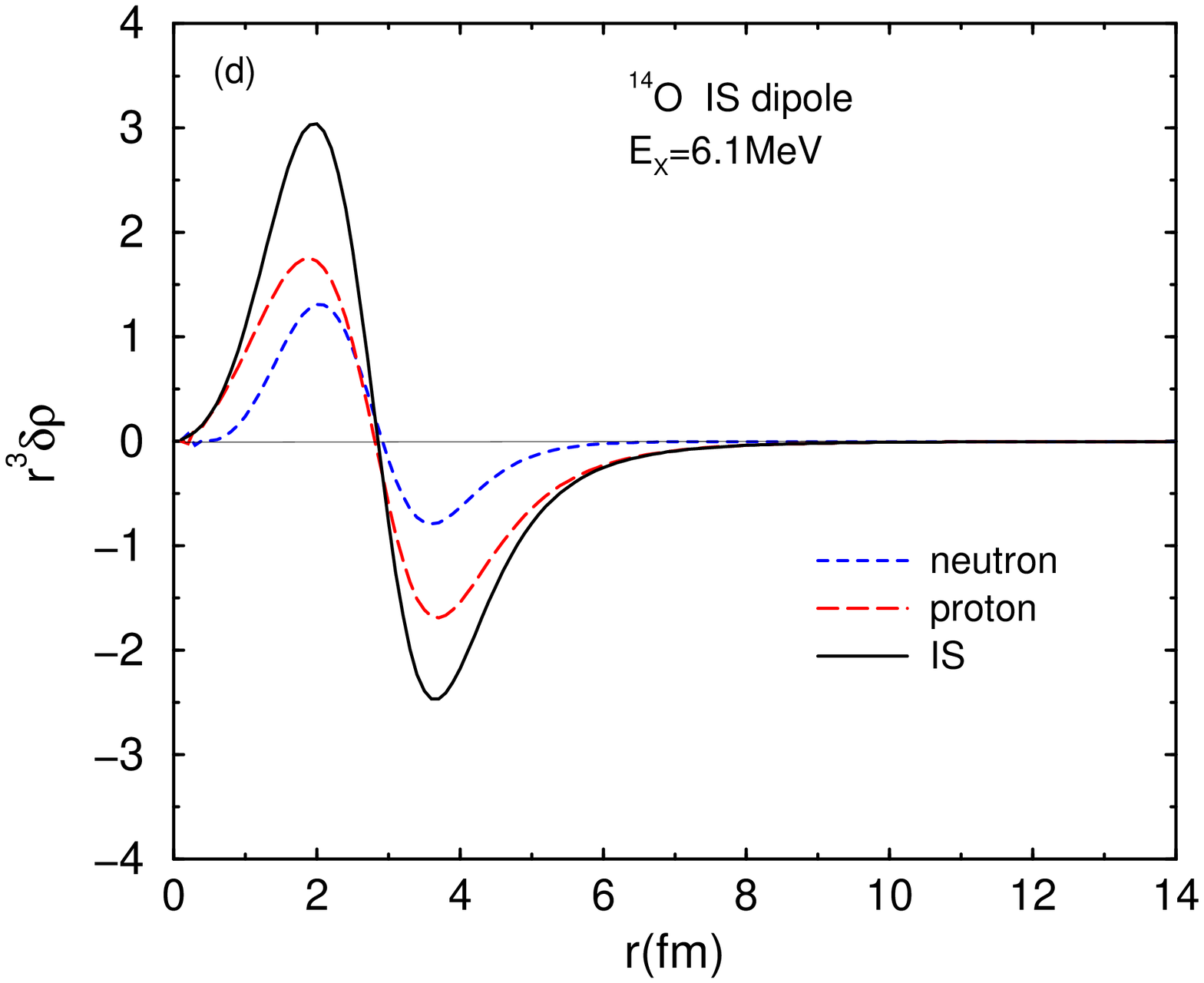}
\vspace{-2cm}
\includegraphics[width=7cm,clip]{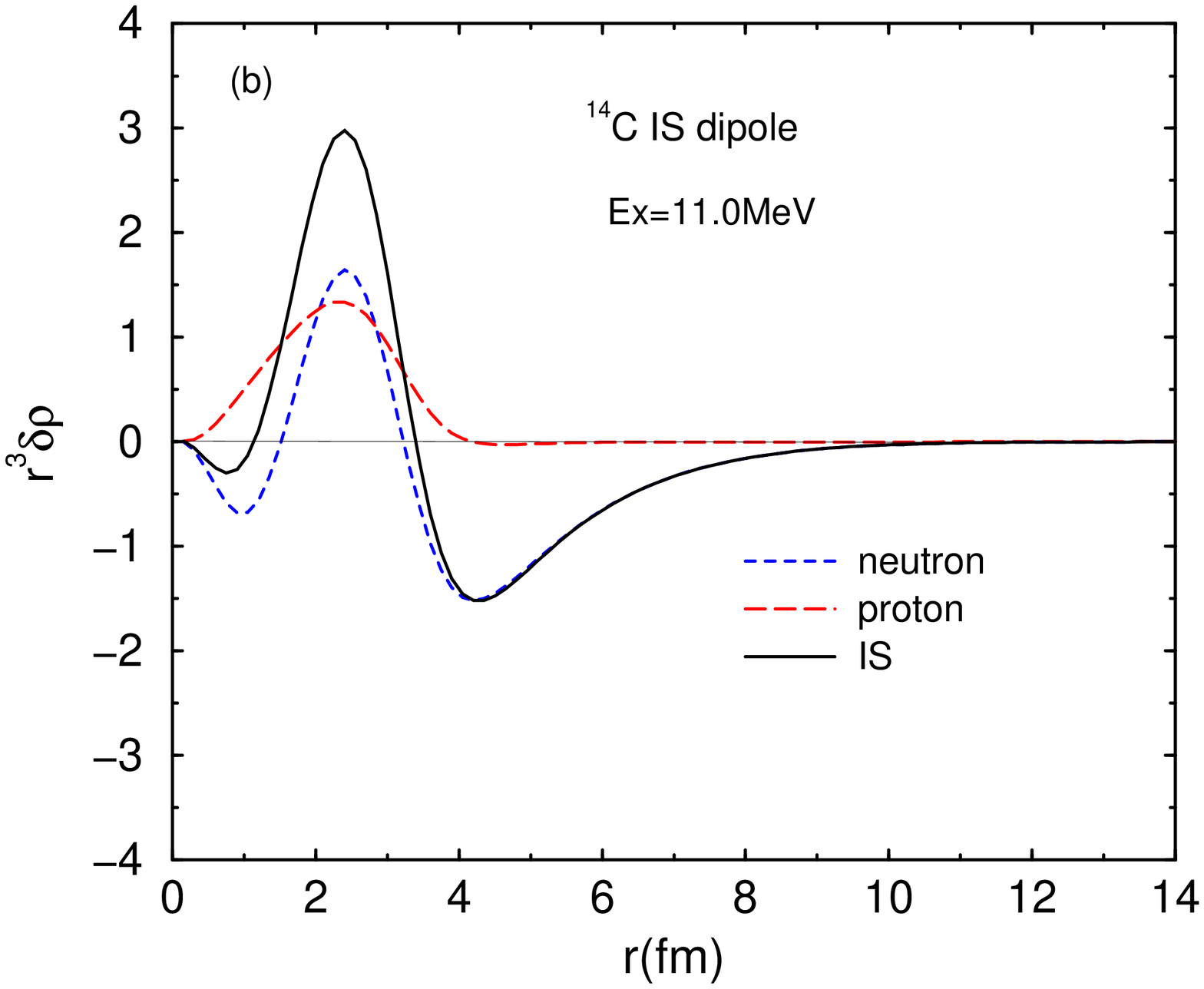}
\includegraphics[width=7cm,clip]{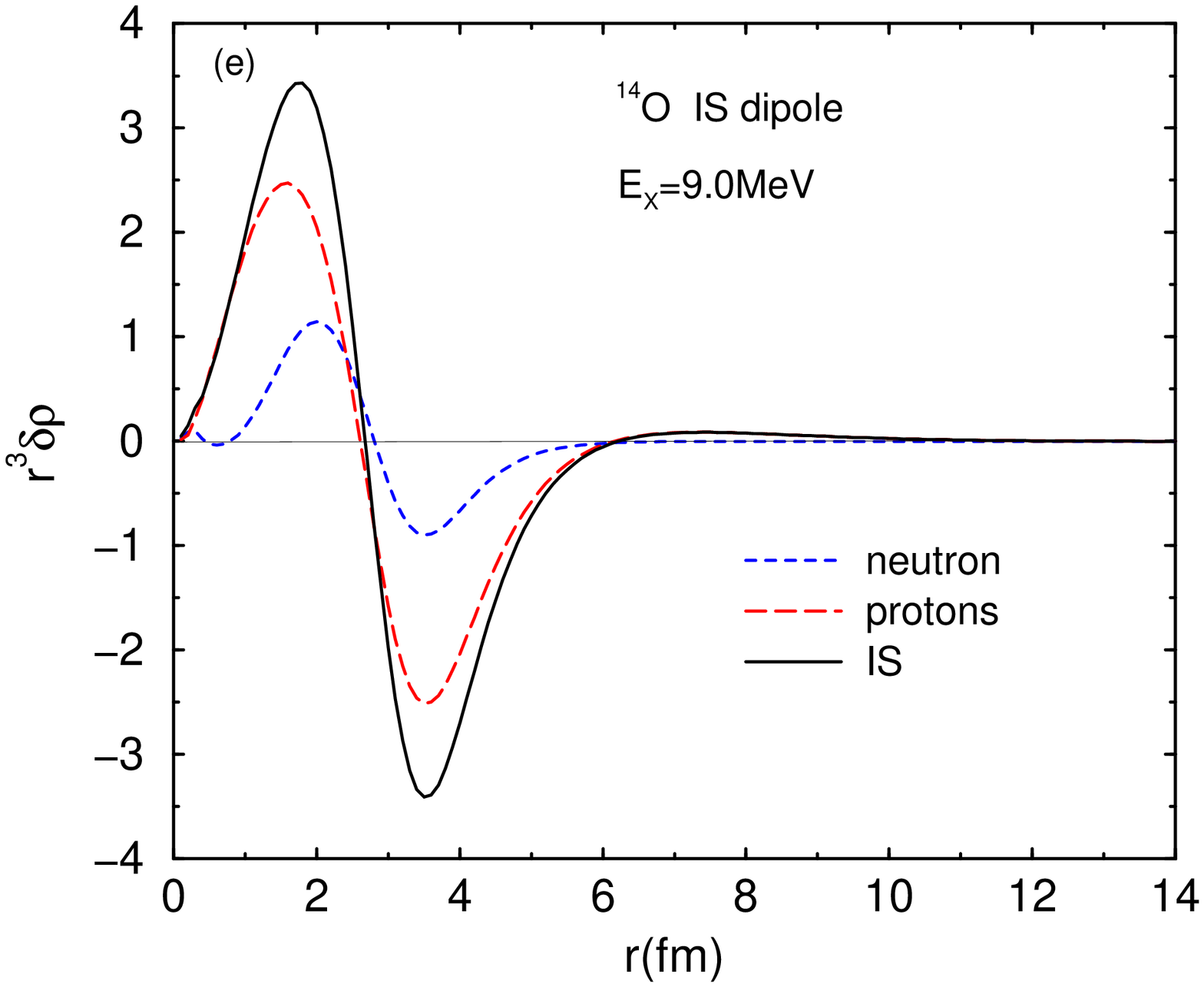}
\includegraphics[width=7cm,clip]{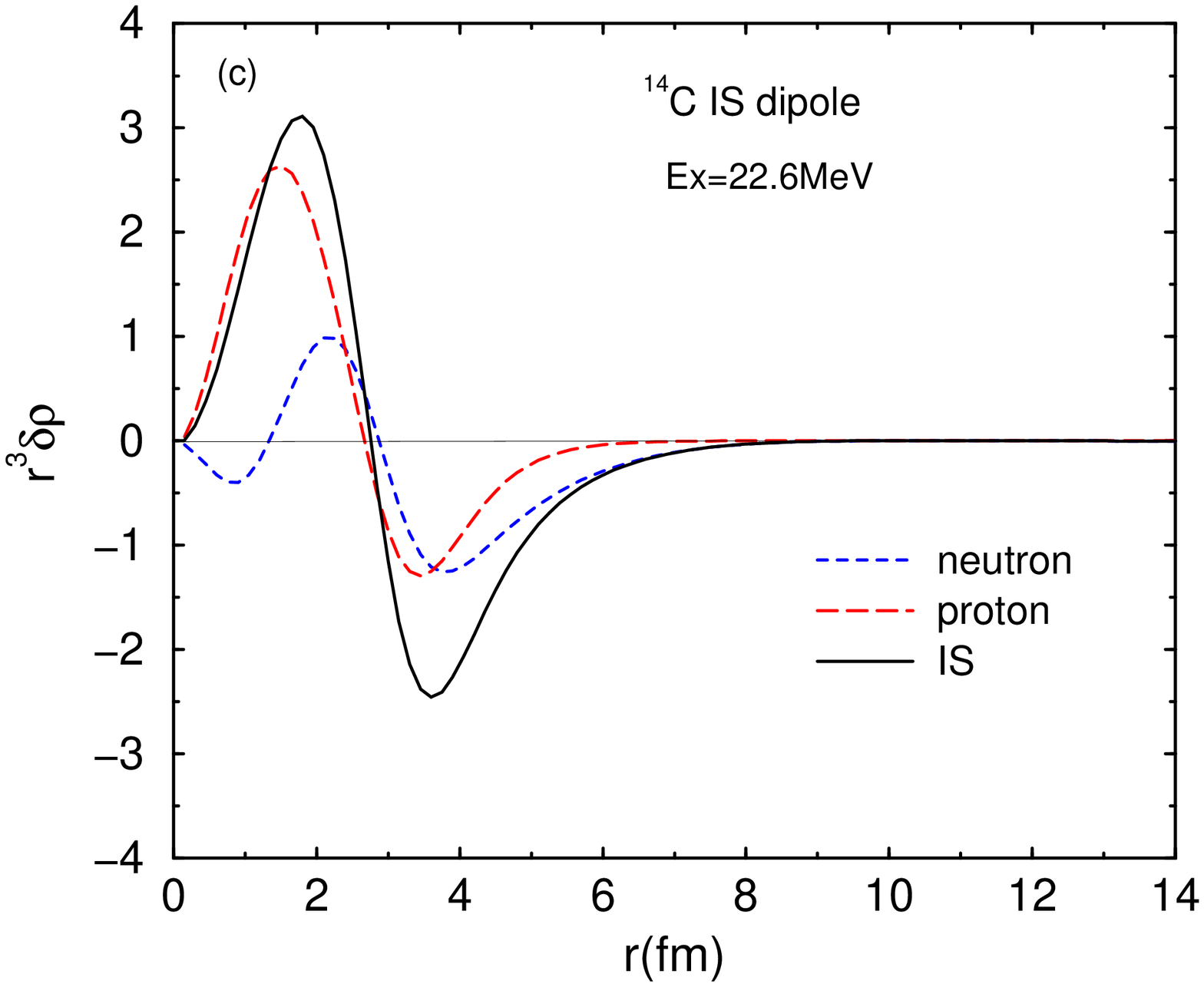}
\includegraphics[width=7cm,clip]{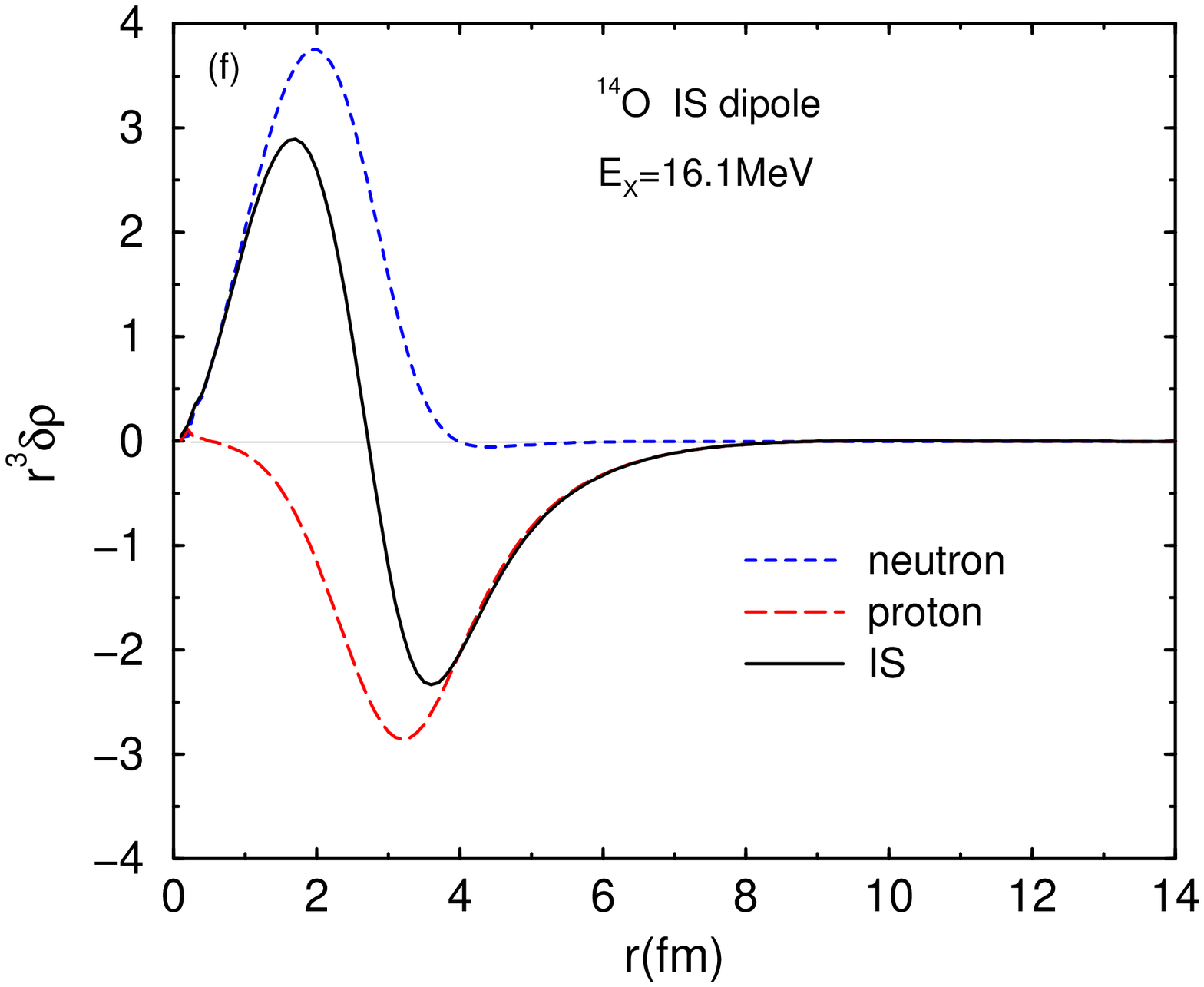}
\caption{\label{fig4}  (Color online) 
Transition densities of the IS dipole states in 
$^{14}$C and $^{14}$O; (a) and (d) of the lowest 1$^-$ states, 
(b) and (e) at the threshold peaks,  and (c) and 
(f) at the GR peaks. The transition density is shown in an arbitrary unit.}
\end{figure}

The transition densities of several  1$^-$ states are shown in Fig. \ref{fig4}; (a) and (d) for the lowest 1$^-$ states, (b) and (e) for 
the threshold peaks,  and (c) and (f) for the GR peaks. 
In order to subtract the spurious component from the transition density,
 we define a modified transition density 
\begin{equation}
\delta \rho -\alpha\frac{d\rho_0}{dr}
\end{equation}
where $\rho_0$ is the total density of the ground state.
The factor $\alpha$ is determined to satisfy  the condition
\begin{equation}
\int(\delta \rho -\alpha\frac{d\rho_0}{dr})r^3dr=0
\end{equation}
which is equivalent to evaluate the strength function with the operator
(\ref{eq:opedpism})\cite{HSZ98}.  
The transition
densities for the lowest  1$^-$ states show  similar radial dependence 
in  Figs. 
\ref{fig4} (a) and (d) for $^{14}$C 
and  $^{14}$O ,respectively, although  the larger contribution 
comes from neutrons in  $^{14}$C and  from 
protons in $^{14}$O.  The transition density of the
peak around Ex=11.0MeV in  $^{14}$C 
 (Fig. \ref{fig4} (b)) is dominated by a large neutron contribution
$\nu$(1p$_{3/2}\rightarrow$2s$_{1/2}$) 
 with a long tail extended until r=10fm, while that of the peak around 
Ex=9.0MeV in $^{14}$O (Fig.  \ref{fig4} (e)) with 
the dominant configuration $\pi$(1p$_{3/2}\rightarrow$s$_{1/2}$) 
does not show 
any long tail.  This difference might be caused by 
  the Coulomb barrier for the 
 dominant 
proton wave functions in  $^{14}$O. 
 It is interesting to notice in 
Fig.  \ref{fig4} (f) that the proton and the neutron contributions have 
no node in the transition densities, but the IS one shows the node 
because of the different extensions of the radial wave functions.  

In ref.\cite{Baba}, the IS dipole strength  in  $^{14}$O
  is observed both in the 
low energy region and the high energy GR region.  The experimental data 
show  two peaks  in the low energy region below 
Ex=10MeV. The integrated strength  exhausts
 about 24\% of the EWSR value below Ex=13.8MeV. 
  The two peak structure and the similar sum rule value (18.5\% of EWSR 
up to 13.8MeV) are found
  in the calculated results of  Fig. \ref{fig3}.  
 On the other hand, 
the observed dipole strength up to GR energy region 
   (up to Ex=25MeV)  amounts to be 
70\% of the EWSR value which is much larger than the calculated 
value of 40.5\%.  

\begin{figure}[p]
\vspace{-2cm}
\includegraphics[width=9cm,clip]{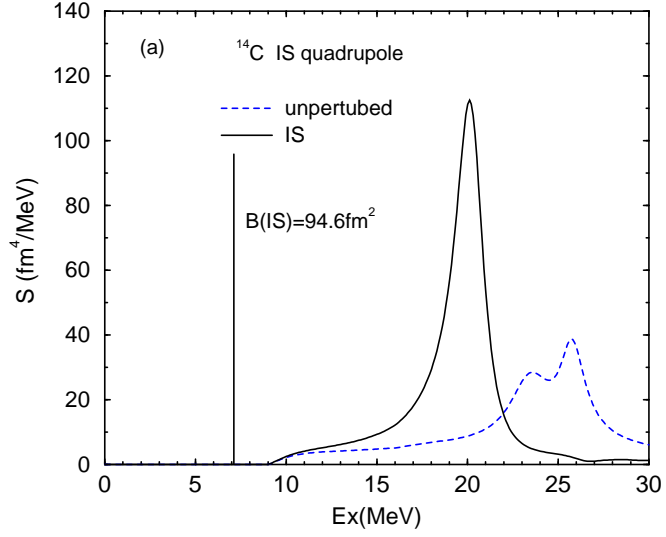}
\includegraphics[width=9cm,clip]{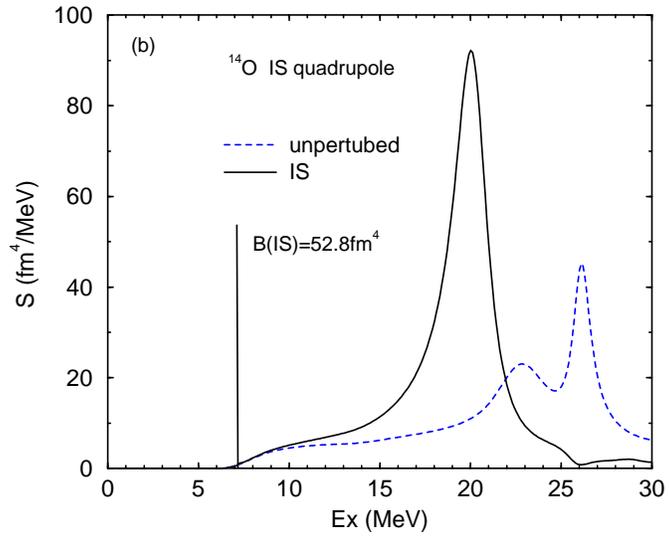}
\caption{\label{fig5} (Color online) 
 IS quadrupole 
 response $S$ in  $^{14}$C and $^{14}$O calculated by
the self-consistent RPA response function theory with the 
Skyrme interaction SkM$^*$.  The operator (\ref{eq:opeqpis}) is used 
in the calculations. 
The solid lines show a discrete  2$^{+}$ state in  $^{14}$C and a narrow 
resonance 2$^{+}$ state in $^{14}$O.  
 See the text for details.}
\end{figure}

In the isoscalar quadrupole  response of 
 Figs. \ref{fig5} (a) and (b), the discrete states are found 
in both nuclei at   Ex=7.1MeV with B(IS)=94.6 fm$^4$ in  $^{14}$C 
and  at Ex=7.2MeV with B(IS)=52.8 fm$^4$ in  $^{14}$O, 
  respectively.  
As far as the excitation energy is concerned, the isospin asymmetry is
much smaller in the quadrupole state than  the dipole states. 
This may be traced back to the difference of 
the particle configurations  between 
 the 1$^-_1$ and 2$^+_1$ states in the
two nuclei.  For the 2$^+_1$ states, the main configurations are 
$\nu$(1p$_{3/2}\rightarrow$1p$_{1/2}$) for  $^{14}$O and 
$\pi$(1p$_{3/2}\rightarrow$1p$_{1/2}$) for $^{14}$C, respectively, and
the  1p$_{1/2}$ states are bound in both nuclei.  
The bound particle states for  for 2$^+_1$ excitations cause only a small
energy difference between  $^{14}$O and $^{14}$C. On the other hand, 
the main configurations for the 1$^-_1$ states are 
$\pi$(1p$_{1/2}\rightarrow$s$_{1/2}$) for  $^{14}$O and 
 $\nu$(1p$_{1/2}\rightarrow$2s$_{1/2}$) for $^{14}$C, respectively.
 The particle state  $\nu$(2s$_{1/2}$) is bound in  $^{14}$C, while 
 the state  $\pi$(s$_{1/2}$) in 
 $^{14}$O is the continuum state.  This difference in 
 the particle configurations
 is the origin of the large difference in the excitation energies of 
 1$^-_1$ states between  $^{14}$O and $^{14}$C.  
 The experimentally observed lowest 2$^+$ state is at Ex=6.59MeV 
in   $^{14}$O and at Ex=7.01MeV in  $^{14}$C.  These data
 are  consistent with the calculated 
smaller isospin asymmetry in the 2$^+$ states  than that  of 
  the  lowest  1$^-$ states.  
Namely,  the observed energy differences  are 
 $\Delta$E=0.92MeV for the  1$^-$ states, 
  while that is  $\Delta$E=0.42MeV in the case of 2$^+$ states. 
 The calculated results show also the same trend.

\begin{figure}[p]
\includegraphics[width=8cm,clip]{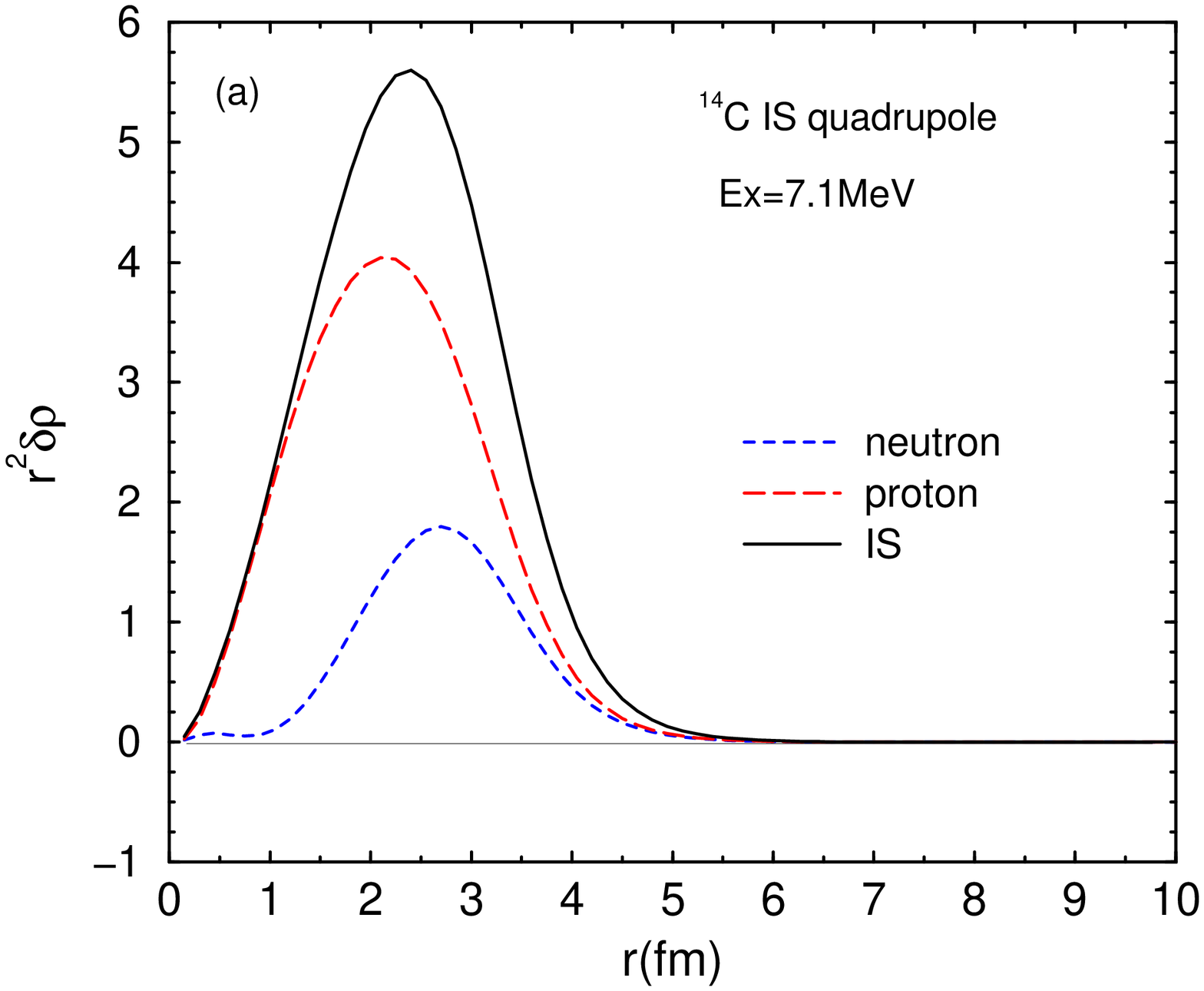}
\includegraphics[width=8cm,clip]{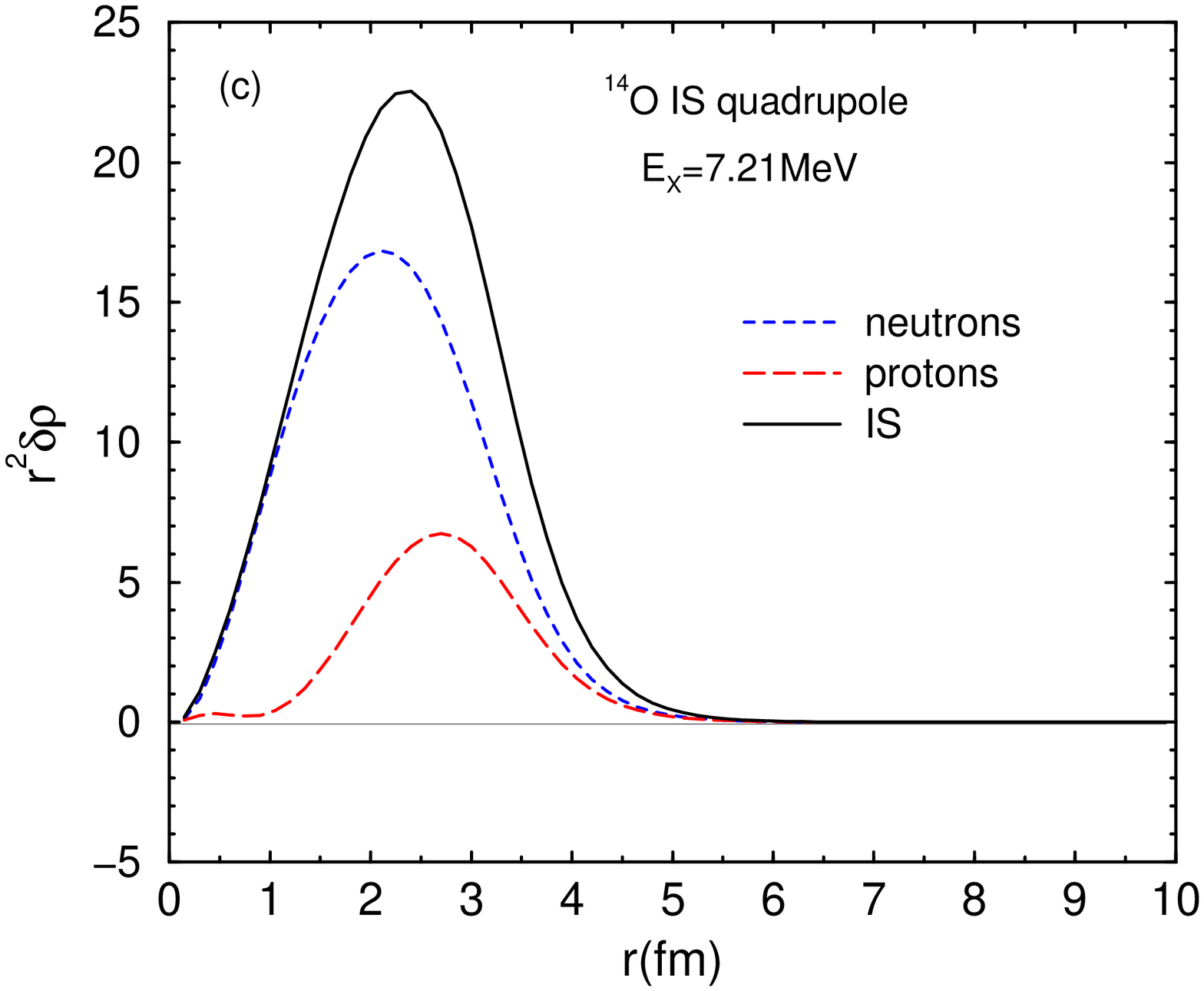}
\includegraphics[width=8cm,clip]{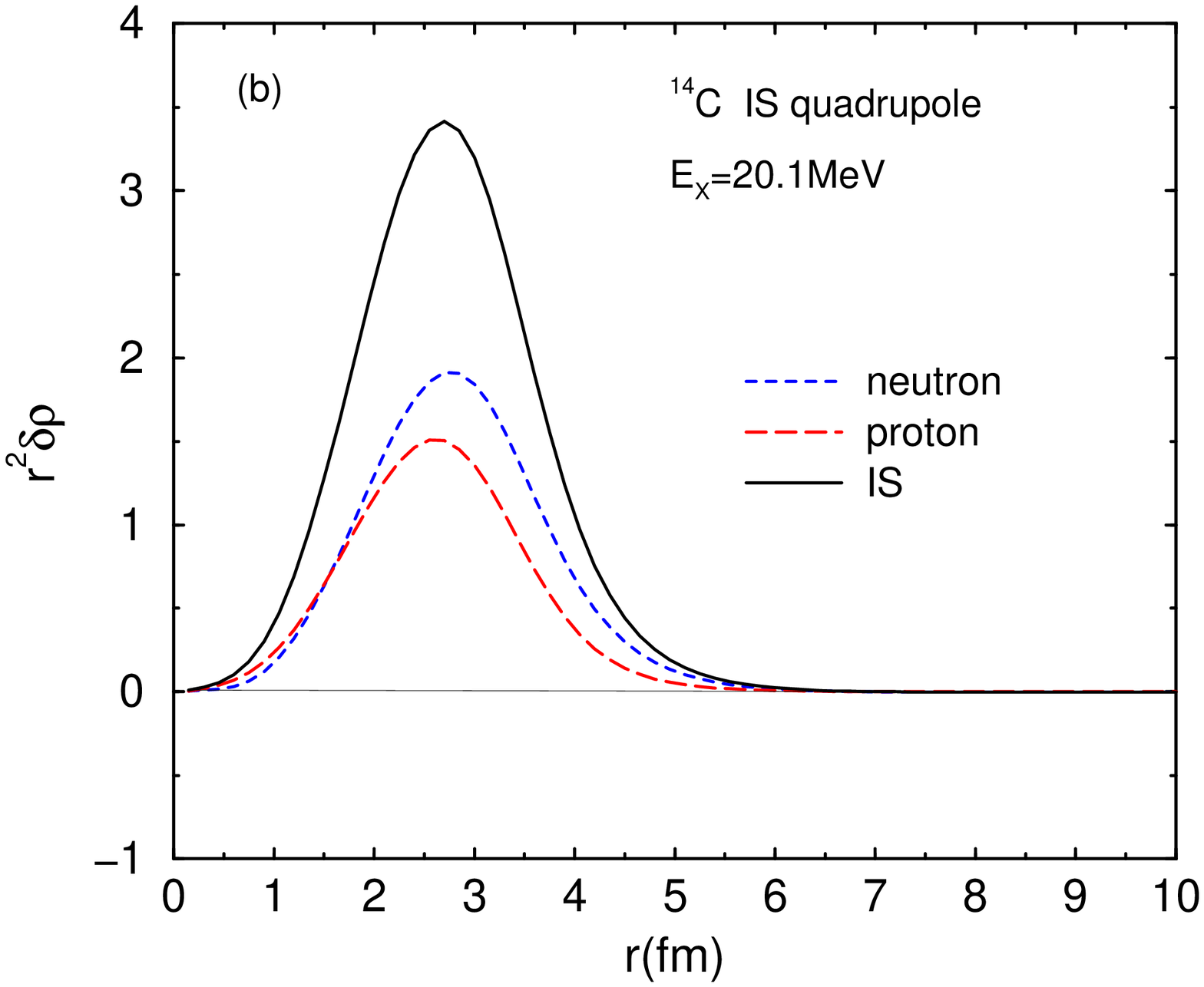}
\includegraphics[width=8cm,clip]{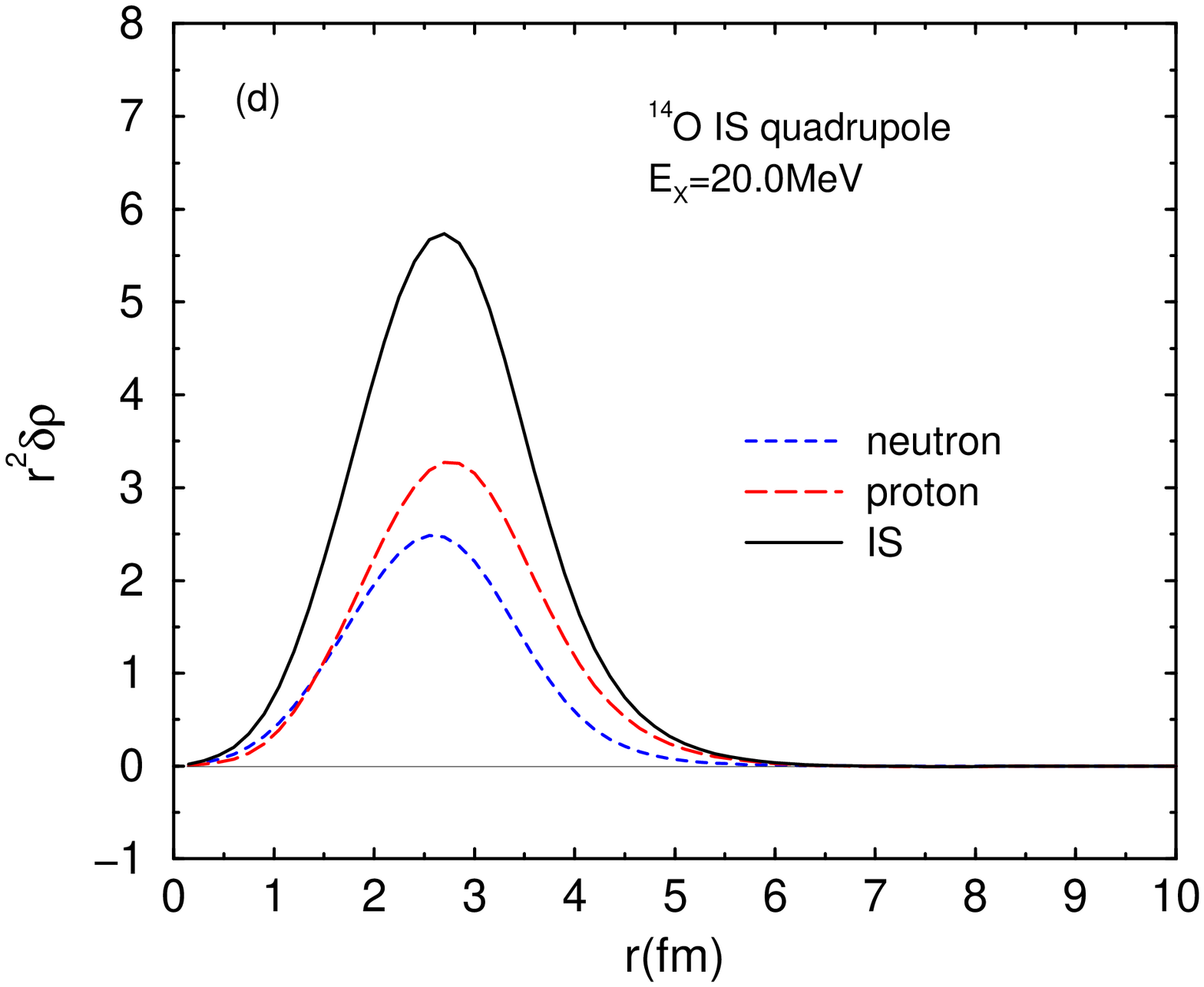}
\caption{\label{fig6} (Color online) 
Transition densities of the IS quadrupole states in 
$^{14}$C and $^{14}$O; (a) and (c) of the lowest 2$^+$ states, 
and (b) and (d) at the GR peaks. The transition density
 is shown in an arbitrary unit.}
\end{figure}

 As is seen in Fig. \ref{fig5}, 
  the threshold strength does not show any  
substantial enhancement in the quadrupole response, 
 but the strong GR peaks are clearly seen around Ex=20MeV 
in the two nuclei.  The EWSR of IS quadrupole response below Ex
=25MeV exhausts 
85.3\% in  $^{14}$C and 85.5\% in $^{14}$O.  As far as the GR is 
concerned, the two nuclei show almost the same excitation spectra both 
in energy and in strength distribution. The transition densities of 
the lowest 2$^+$  and GR states are shown in Fig.  \ref{fig6}. 
It is interesting to notice that the proton 
amplitude is larger in the lowest 2$^+$ state  of
  $^{14}$C, while the neutrons have the dominant 
contribution in $^{14}$O.  This is because  the proton  excitation
 (1p$_{3/2}\rightarrow$1p$_{1/2}$)
 is allowed in  $^{14}$C, while   the neutron excitation
 (1p$_{3/2}\rightarrow$1p$_{1/2}$)  is allowed  in  $^{14}$O.
 In   Figs. \ref{fig6} (b) and (d), 
the transition densities  of quadrupole 
  GR show  a typical collective IS nature, i.e., 
   the radial dependence is  surface peaked, so called,
  Tassie-type  and  the neutron (proton) 
contribution is larger than  the proton (neutron) one  in the neutron-rich 
(proton-rich) nucleus  $^{14}$C ($^{14}$O).  One can not see any substantial 
long-tail in the transition densities of the quadrupole response even in the 
low energy states.  This is due to a larger centrifugal barrier for the
 quadrupole response than  the monopole and the dipole responses
 \cite{Sagawa92}.

\section{SUMMARY}
We studied the isospin asymmetry in the IS 
  multipole responses of A=14 mirror
 nuclei  $^{14}$C and  $^{14}$O  using the  HF+RPA theory with 
  Skyrme interaction
to take into account the continuum coupling properly.  
We found that the IS strength distributions near the threshold show a large 
 asymmetry between the two nuclei for the monopole and the dipole responses.
On the other hand, the strength distributions in the GR region are very
 similar  in both the peak energy and the sum rule value.
It is seen a clear sign of the halo-type extension of the transition densities 
in  the threshold 
 monopole response. Contrarily,  those of GR peaks do  not show any sign of the
halo effect, but give  a typical IS collective features. For the quadrupole 
response, the two nuclei show very similar features not only in the GR region, 
but also in the lower energy region near the threshold. 
The calculated results of  $^{14}$O are compared with recent experimental data 
obtained by the multipole 
decomposition analysis of $\alpha$ inelastic scattering.  The calculated 
 monopole and the dipole strength distributions  show good agreement  with
the empirical ones near the threshold.  The empirical sum rule strength 
of the monopole states is also close to the calculated one in the GR region 
below Ex=25MeV, while the empirical dipole strength is
reported  to be 
larger than the calculated one in the high energy region between 
12MeV$<$ Ex $<$25MeV.  In order to confirm further the isospin 
asymmetry of A=14 nuclei, we need  the experimental data 
of $^{14}$C in the continuum region.

\begin{acknowledgments}
I thank  T. Shimoura for enlightening discussions.  I thank also
 H. Baba for informing his data prior 
to a publication.  This work is supported in part by the Japanese
Ministry of Education, Culture ,Sports, Science  and Technology
  by Grant-in-Aid  for Scientific Research under
 the program number (C(2)) 16540259.
\end{acknowledgments}

\newpage
\noindent {\bf\large Figures }

\newpage

\newpage

\newpage

\newpage

\begin{thebibliography}{99}
\bibitem{Heisenberg32} W. Heisenberg, Z. Physik {\bf 77}, 1 (1932).
\bibitem{BM69}  A. Bohr and B. R. Mottelson, Nuclear Structure, Vol. I
 (Benjamin, New York, 1969). 
\bibitem{HS93}
 I. Hamamoto and  H. Sagawa,  Phys. Rev. C{\bf 48}, R960 (1998).
\bibitem{HGS95}
H. Sagawa, Nguyen van Giai and T. Suzuki, Phys. Lett.  {\bf B353}, 7 (1995).
\bibitem{Fox64} J. D. Fox, C. F. Moore and D. Robson, Phys. Rev. Lett. {\bf 12}, 1981 (1964).
\bibitem{ET51}  J. B. Ehman, Phys. Rev. {\bf 81}, 412 (1951). \\
R. G. Thomas,  Phys. Rev. {\bf 88}, 1109 (1952).
\bibitem{Matsuta95}
 K. Matsuta et al., Nucl. Phys. {\bf A588}, 153c (1995);  K. Matsuta et al., 
 Hyperfine Interact. {\bf 97/98}, 519 (1996).  
\bibitem{Utsuno04}
  Y. Utsuno,  Phys. Rev. C{\bf 70}, 0110303R(2004).
\bibitem{Moto91}
 P. Decrock et al., Phys. Rev. Lett. {\bf 67}, 808(1991).\\
 T. Motobayashi et al., Phys. Lett. {\bf B264}, 259(1991).
\bibitem{BT75}  G. F. Bertsch and S. F. Tsai, Phys. Report {\bf 18}, 
125 (1975).
\bibitem{LG76}  S. Shlomo and  G. F. Bertsch, Nucl. Phys. {\bf A243}, 507 (1975);\\
S. F. Liu and Nguyen van Giai, Phys. Lett.  {\bf 65B}.  
23 (1976). 
\bibitem{Harakh81}
 M. N. Harakeh and A. E. L. Dieperink, Phys. Rev. C{\bf 23}, 2329(1981).\\
 Nguyen Van Giai and H. Sagawa, Nucl. Phys. {\bf A371}, 1 (1981). 
\bibitem{HSZ98}
I. Hamamoto, H. Sagawa and X. Z. Zhang,  Phys. Rev. C{\bf 57}, R1064 (1998).
\bibitem{Khan}
E. Khan, N. Sandulescu, M. Grasso and Nguyen Van Giai, Phys. Rev. 
C{\bf 66}, 024309 (2002). 
\bibitem{Colo}
 C. Mahaux, P.F. Bortignon, R.A. Broglia and C. H. Dasso, Physics Reports
 120,1 (1985). \\
 G. Col\`o, Toshio Suzuki and H. Sagawa,  Nucl. Phys. {\bf A695}, 167 (2001).
\bibitem{HSZ97}
I. Hamamoto, H. Sagawa and X. Z. Zhang,  Phys. Rev. C{\bf 55}, 2361 (1997).
\bibitem{Sagawa02}
H. Sagawa, Phys. Rev. C{\bf 65}, 064314(2002).
\bibitem{Baba}
  H. Baba, Doctoral Thesis (2005);
H. Baba et al., RIKEN Accel. Prog. Rep. {\bf 38}, 48(2005) and
private communications.
\bibitem{HS02}
I. Hamamoto and H. Sagawa, Phys. Rev. C{\bf 66}, 044315(2002).
\bibitem{Firestone}
 Table of Isotopes I (edited by R. B. Firestone, 1996,
  John Wiley and Sons, Inc)
\bibitem{Sagawa92}
 H. Sagawa, Phys. Lett.  {\bf B286}, 7(1992).\\
 D. V. Fedorov, A. S. Jensen and K. Riisager, Phys. Lett.  
 {\bf B312}, 1(1993). 
\end{thebibliography}
\end{document}